\documentclass[aps,prl,10pt,twocolumn,showpacs, superscriptaddress]{revtex4-1}

\usepackage{graphicx}
\usepackage{dcolumn}   
\usepackage{bm}        
\usepackage{multirow}
\usepackage{amssymb}
\usepackage{amsfonts}   
\usepackage{dsfont}	
\usepackage{amsmath} 
\usepackage{amsthm}
\usepackage{color}
\usepackage{marvosym}

\usepackage[normalem]{ulem}

\newcommand{\mathbbm}[1]{\text{\usefont{U}{bbm}{m}{n}#1}} 
\usepackage[utf8]{inputenc}

\newcommand{\tr}{\text{tr}}
\newcommand{\one}[0]{\mathds{1}}		
\newcommand{\ot}[0]{\otimes}
\newcommand{\nn}[0]{\nonumber}

\newcommand{\dyad}[1]{| #1\rangle \langle #1|}  
\newcommand{\nhf}[0]{\lfloor n/2 \rfloor}
\newcommand{\supp}[0]{\operatorname{supp}}

\newcommand{\ket}[1]{|#1\rangle}

\newcommand{\R}{\ensuremath{\mathbbm R}}
\newcommand{\C}{\ensuremath{\mathbbm C}}
\newcommand{\be}{\begin{equation}}
\newcommand{\ee}{\end{equation}}
\newcommand{\bea}{\begin{eqnarray}}
\newcommand{\eea}{\end{eqnarray}}

\newcommand{\kommentar}[1]{}

\newcommand{\forget}[1]{}

\begin{document}

\title{Absolutely maximally entangled states of seven qubits do not exist}
\date{\today}

\author{Felix Huber}
\affiliation{Naturwissenschaftlich-Technische Fakult\"at, Universit\"at Siegen, 
57068 Siegen, Germany}
\author{Otfried G\"uhne}
\affiliation{Naturwissenschaftlich-Technische Fakult\"at, Universit\"at Siegen, 
57068 Siegen, Germany}
\author{Jens Siewert}        
\affiliation{Departamento de Qu\'{i}mica F\'{i}sica, Universidad del Pa\'{i}s Vasco UPV/EHU, E-48080 Bilbao, Spain}
\affiliation{IKERBASQUE Basque Foundation for Science, E-48013 Bilbao, Spain}

\begin{abstract}
Pure multiparticle quantum states are called absolutely maximally entangled 
if all reduced states obtained by tracing out at least half of the particles
are maximally mixed. We provide a method to characterize these states for a 
general multiparticle system. With that, we prove that a seven-qubit state 
whose three-body marginals are all maximally mixed, or equivalently, a pure 
\( ((7,1,4))_2 \) quantum error correcting code, does not exist. Furthermore, 
we obtain an upper limit on the possible number of maximally mixed three-body 
marginals and identify the state saturating the bound. This solves the 
seven-particle problem as the last open case concerning maximally entangled 
states of qubits.
\end{abstract}

\maketitle

{\it Introduction.---}
Multiparticle entanglement is central for the understanding of the possible
quantum advantages in metrology or information processing. When investigating 
multiparticle entanglement as a resource, the question arises which quantum states
are most entangled. For a pure multiparticle quantum state maximal entanglement 
is present across a bipartition if the smaller of the two corresponding reduced 
systems is maximally mixed. It is then a natural question to ask whether or not 
there exist quantum states for any number of parties \(n\), such 
that all of its reductions to \(\lfloor n/2 \rfloor\) parties 
are maximally mixed \cite{Gisin1998, Higuchi2000}. If this is the 
case, maximal entanglement is present across all bipartitions and, 
accordingly, these states are also known as {\em absolutely maximally 
entangled} (AME) states 
\cite{Brown2005, Helwig2012, Arnoud2013, Kloeckl2015, Goyeneche2015, Helwig2013_graph, Bernal2016, Borras2007, Facchi2008,
Facchi2010, Gour2010, Zyczkowski2016, Zha2011, Helwig2013, Scott2004, Chen2016}.
These states have been shown to be a resource for open-destination and 
parallel teleportation \cite{Helwig2012}, for threshold quantum secret 
sharing schemes \cite{Helwig2013}, and are a type of quantum error correcting codes \cite{Scott2004}.

If the local dimension is chosen large enough, AME states always exist 
\cite{Helwig2013}. For qubits, however, the situation is only partially resolved. 
The three-qubit Greenberger-Horne-Zeilinger (GHZ) state is an AME state 
since all the single-qubit reduced states are maximally mixed. 
For four qubits it was shown that AME states do not exist
\cite{Higuchi2000} and best approximations of AME states 
(where not all reduced states are maximally mixed) have been presented
\cite{Gour2010}. Five- and six-qubit AME states are known 
\cite{Scott2004, Borras2007, Facchi2010}. These can be represented as 
graph states and correspond to additive or 
stabilizer codes used in quantum error correction \cite{Hein2006, Scott2004}. 
For more than eight qubits, AME states do not exist 
\cite{Scott2004, Rains1998, Rains1999, Nebe2006}.

Despite many attempts, the case of seven qubits remained unresolved. 
Numerical results give some evidence for the absence of an AME state \cite{Borras2007, Facchi2008,Facchi2010}. 
By exhaustive search, it was shown that such a state could not have the form 
of a stabilizer state \cite{Hein2006}. Nevertheless, some approximation has 
been presented by making many but not all three-body marginals maximally 
mixed~\cite{Zha2011, Goyeneche2014}.

As shortly mentioned, AME states are a type of pure {\em quantum error correcting codes} (QECC),
having the maximal distance allowed by the Singleton bound \cite{Grassl2015}.
In particular, AME states of \(n\) parties having local dimension \(D\) each
correspond to a pure QECC in \((\C^D)^{\ot n}\) of distance \(\nhf+1\), 
denoted by \(((n,1,\nhf+1))_D\).
Often, but not always, bounds on so-called non-additive (i.e. non-stabilizer) codes 
coincide with those for additive (stabilizer) codes.
The seven qubit AME state would - if it existed - be one of the few examples
where a non-additive code outperformed an additive one.
This possibility was noted already in a seminal article by Calderbank et al. \cite{Calderbank1998}.
Up to \(n=30\), there are only three other instances known where this could
be the case for one-dimensional codes on qubits \cite{open_cases}.

In this paper, we provide a method to characterize AME states and their 
approximations, making use of the Bloch representation~\cite{Fano1957}. The usefulness 
of this tool may be surprising at first sight, as the Bloch representation is
designed to be a tool for mixed states. We were motivated to choose this 
approach by the fact that monogamy equalities~\cite{Gour2010,Eltschka2015} directly signal the non-existence 
of a four-qubit AME state, and the natural framework 
for deriving the monogamy equalities appears to be the Bloch representation 
\cite{Eltschka2015}.

{\it The Bloch representation.---}
Any \(n\)-qubit state can be written as
\begin{equation}
	\varrho = \frac{1}{2^n} 
	\sum_{\alpha_1 \dots \alpha_n} r_{\alpha_1, \dots, \alpha_n} \sigma_{\alpha_1} \ot \cdots \ot \sigma_{\alpha_n} \,,
\end{equation}
where the  \(\{\alpha_1, \dots, \alpha_n \} \in \{0, x,y,z\}\) label combinations 
of the four Pauli matrices.
For simplicity, we group the terms according to their weight, 
that is, their number of non-trivial (Pauli) operators.
Let \(P_j\) be the sum over terms of weight \(j\), then the state can be written as
\begin{equation}
	\varrho = \frac{1}{2^n}\Big(\one^{\ot n} + \sum_{j=1}^n P_j\Big)\,.
\end{equation}
We denote by \(P_j^{(V)}\) a subset of \(P_j\), where 
$V$ further specifies its support, i.e. its non-trivial 
terms are located  on the subsystems in \(V\). To give an
example, a state of three qubits reads
\begin{equation}
	\varrho = \frac{1}{2^3}\Big(\one^{\ot 3} + \sum_{j=1}^3 P_1^{(j)} + \sum_{1 \leq k < l \leq 3} P_2^{(kl)} + P_3\Big)\,,
\end{equation}
where, e.g., \(P_2^{(12)} = \sum r_{\alpha_1,\alpha_2, 0} \,\sigma_{\alpha_1} \ot \sigma_{\alpha_2} \ot \one \)
and \(\alpha_1, \alpha_2 \neq 0\).
When tracing out the third qubit, one drops the terms \(P_3, P_2^{(13)}, P_2^{(23)},\)
and \(P_1^{(3)}\), as they do not contain an identity in the third subsystem. Also, the 
normalization prefactor is multiplied by the dimension of the parties over which the partial trace was performed, resulting in
\begin{equation}
\label{eq:3qb_trace}
	\tr_{\{3\}} [\varrho] \ot \one = \frac{1}{2^2}
	\big(\one^{\ot 3} + P_1^{(1)} + P_1^{(2)} + P_2^{(12)}\big)\,.
\end{equation}
Accordingly, a three-qubit state having maximally mixed one-body reduced density matrices 
does not have terms of weight one, the terms \(P_1^{(j)}\) are absent. 
Similarly, in \(n\)-qubit AME states all operators \(P_j\) with \(1 \leq j \leq \nhf\) vanish.

Our further discussion rests on recognizing what terms may appear in 
the squared state \(\varrho^2\). For this, consider two terms \(A\) and \(B\), 
both appearing in the Bloch expansion of the state. For computing \(\varrho^2\), 
the anticommutator \(\{A,B\}\) is required, and we state the following observation regarding 
its weight.

\noindent
{\bf Lemma 1 (parity rule).}
Let \(M, N\) be Hermitian operators proportional to \(n\)-fold 
tensor products of single-qubit Pauli operators,  
\(M = c_M \,\sigma_{\mu_1} \ot\cdots \ot \sigma_{\mu_n}\), 
\(N = c_N \,\sigma_{\nu_1} \ot\cdots \ot \sigma_{\nu_n}\), 
where \(c_M, c_N \in \R\).
Let us denote their weights, that is, their number of nontrivial Pauli-operators in 
their tensor expansion, by \(|M|\) and \(|N|\). 
Then, if the anticommutator $\{M,N\}$ does not vanish, 
its weight \(|\{M,N\}|\) fulfills
\begin{equation}
	|\{M,N\}| = |M| + |N| \mod 2\,.
\end{equation}
\begin{proof}
  The product \(MN\), and thus also \(\{M,N\}\), has at most weight 
  \(|M|+|N|\). This is attained, if the supports of $M$ and $N$ are disjoint. 
  Each pair of equal, but non-zero indices \(\mu_j = \nu_j\) corresponds to some 
  overlap of the supports and reduces the maximal weight \(|M|+|N|\) by two. 
  In contrast, if a pair of   non-zero indices are not equal (e.g., 
  \(\mu_j \neq \nu_j\)), the product \(MN\) contains the 
  term \(\sigma_{\mu_j} \sigma_{\nu_j} = i \epsilon_{\mu_j \nu_j \chi} \sigma_\chi\).  
  Consequently for each such pair \(|M|+|N|\) is reduced by only one. 
  If an odd number of such pairs exists, the anticommutator has to vanish, as it is Hermitian. 
  So, such pairs have to occur an even number of times, which proves the claim. 
\end{proof}

We can summarize the behavior of the weights of \(M\) and \(N\) and 
their anticommutator as follows:
\begin{align}
  \{ \rm{even}, \rm{even} \} \quad &\longrightarrow \quad \rm{even} , \nn\\
  \{ \rm{odd}, \rm{odd} \}   \quad &\longrightarrow \quad \rm{even} , \nn\\
  \{ \rm{even}, \rm{odd} \}  \quad &\longrightarrow \quad \rm{odd} .
\end{align}
It follows that an analogous behavior holds for the $P_j$. If 
$j$ and $k$ are either both even or both odd, the anticommutator
$\{P_j, P_k\}$ can only contribute to the $P_l$ where $l$ is
even. Similarly, if $j$ is even and $k$ is odd, it only contributes
to the $P_l$ having odd $l$.

{\it Properties of AME state reductions.---}
First, recall that for a pure \(n\)-party state \(\ket{\psi}_{AB}\) 
consisting of \(D\)-level systems, 
the complementary reduced states of any bipartition share the same spectrum. This follows from
its Schmidt decomposition. Hence, if a \((n-k)\)-body 
reduction \(\varrho_B\) is maximally mixed, its complementary reduced 
state \(\varrho_A\) of size \(k\geq \nhf\) has all \(D^{(n-k)}\) nonzero eigenvalues 
equal to \(\lambda = D^{-(n-k)}\).
Thus the reduced state is proportional to a projector,
\begin{equation}\label{eq:projector_property}
    \varrho_A^2 = D^{-(n-k)} \varrho_A\,.
\end{equation}
This projector property alone is already enough to derive the following bounds on the existence of AME states, 
\begin{equation}
  n \leq 
  \begin{cases}
  	2(D^2-1)  &\quad n\,\, \rm{even,} \\
  	2D(D+1)-1 &\quad n\,\, \rm{odd}.
  \end{cases}
\end{equation}
These bounds originate in work by Rains and were applied to AME states 
by Scott \cite{Rains1998, Scott2004}.
A proof using the projector property can be found in Appendix A \cite{suppmat}.

By Schmidt decomposition, one further sees that the full state \(\ket{\psi}_{AB}\)
is an eigenvector of the reduced state \(\varrho_A\),
\begin{equation}\label{eq:eigenvector_property}
	\varrho_A \ot \one^{\ot (n-k)} \ket{\psi}_{AB} = D^{-(n-k)} \ket{\psi}_{AB} \,.
\end{equation}
Accordingly, for an AME state having all \(\nhf\)-body reduced states maximally mixed, 
any \(k\)-body reduced state \(\varrho_{(k)}\) with \(\nhf \leq k \leq n\) fulfills relations 
\eqref{eq:projector_property} and \eqref{eq:eigenvector_property}.

Let us now consider AME states of \(n\) qubits.
We decompose Eq.~\eqref{eq:eigenvector_property} in the Bloch representation, 
using the reduced state \(\varrho_{(k)}\) on the first \(k=\nhf+1\) parties of a qubit AME state,
\begin{equation}
	\frac{1}{2^k}(\one^{\ot k} + P_{k}^{(1\cdots k)}) \ot \one^{(n-k)} \ket{\psi} 
	= 2^{-(n-k)}\ket{\psi}\,.
\end{equation}
Because all \(\nhf\)-body marginals are maximally mixed, \(P_{j \leq \nhf} = 0\). 
We obtain the eigenvector relations
\begin{align}
  P_{k}^{(1\cdots k)} \ot \one^{\ot (n-k)} \ket{\psi} = 
  \begin{cases} 
	  3 \ket{\psi} \quad n\,\, \rm{ even} \,,\\
	  1 \ket{\psi} \quad n\,\, \rm{ odd} \,.
  \end{cases}
\end{align}
By accounting for combinatorial factors, similar relations can be obtained in an 
iterative way for all \(P_{j \geq \nhf+1}^{(1\cdots j)}\).

With these building blocks in place, we are in the
 position to solve the 
last open qubit case --- 
the existence of a seven qubit AME state.
In the following, we will combine the projector property of a five qubit 
reduced state \(\varrho_{(5)}\) with the eigenvector relations 
for \(P_4^{(1\cdots 5)}\) and \(P_5^{(1\cdots 5)}\) 
to obtain a contradiction from the parity rule stated in Lemma 1.

\noindent
{\bf Observation 2.}
{\it Consider a pure state of seven qubits. Then not all of its three-body 
reduced density matrices can be maximally mixed.}

{\it Proof.}
Assume we have a pure seven-qubit state \(\varrho = \dyad{\phi}\), whose three-body 
marginals are all maximally mixed. Then, its five-party 
reduced density matrix on systems \(\{1,\cdots ,5\}\) is proportional to a projector,
\begin{equation}
\label{projequation}
\varrho_{(5)}^2 = \frac{1}{4}\varrho_{(5)}\ .
\end{equation}
Note that while the proof requires the projector property only to hold on the first five qubits, 
Eq.~\eqref{projequation} actually holds for all possible five-qubit reductions.

Regarding the eigenvector relations, a Schmidt decomposition of the pure state $\ket{\phi}$
across the bipartitions \(\{1,2,3,4\,|\,5,6,7\}\) and 
\(\{1,2,3,4,5 \, | \, 6,7\}\) yields
\begin{align}
	\varrho_{(4)} \ot \one^{\ot 3} \,\ket{\phi} &= \frac{1}{8} \ket{\phi} \,,
	\label{evequation1}
	\\
	\varrho_{(5)}\ot \one^{\ot 2}  \,\ket{\phi} &= \frac{1}{4} \ket{\phi} \,.
	\label{evequation2}
\end{align}
Again, analogous equations hold for any possible four- or five-qubit reductions,
including for the five different four-party reduced states in \(\{1,2,3,4,5\}\).

We will use these three equations to obtain a contradiction: 
Let us expand \(\varrho_{(4)}\) and \(\varrho_{(5)}\) in the Bloch basis
\begin{align}
	\varrho_{(4)} &= \frac{1}{2^4} (\one + P_4 )\,,
	\label{blochfourequation}
	\\
	\varrho_{(5)} &=  \frac{1}{2^5} (\one + \sum_{j=1}^5 P_4^{[j]}\ot\one^{(j)} + P_5) \,.
	\label{blochfiveequation}
\end{align}
There are five different terms \(P_4^{[j]}\ot\one^{(j)}\), with \([j]\) indexing 
the five different supports of weight four terms within a five body reduced state, 
each having an identity on different positions.

Inserting Eqs.~(\ref{blochfourequation}, \ref{blochfiveequation}) into 
Eqs.~(\ref{evequation1}, \ref{evequation2}) results in the eigenvector relations
\begin{align}
	 P_4^{[j]}  \ot \one^{\ot 3}  \,\ket{\phi} &= 1 \ket{\phi} \,,\nn\\
	 P_5        \ot \one^{\ot 2}   \,\ket{\phi} &= 2 \ket{\phi} \,.
\end{align}
We similarly insert Eq.~(\ref{blochfiveequation}) in Eq.~(\ref{projequation})
to obtain
\begin{align} \label{eq:rho_5_sq}
	&\big(\one + \sum_{j=1}^5 P_4^{[j]}\ot\one^{(j)} + P_5\big)\big(\one + \sum_{j=1}^5 P_4^{[j]}\ot\one^{(j)} + P_5\big) \nn\\
	&= 8 \big(\one + \sum_{j=1}^5 P_4^{[j]}\ot\one^{(j)} + P_5\big)\,.
\end{align}
The key observation is now the parity rule stated in Lemma~$1$: 
Only certain products occurring on the left-hand side of Eq.~\eqref{eq:rho_5_sq}
can contribute to $P_5$ on the right-hand side. 
Indeed, \(P_5^2\) on the left-hand side cannot contribute to \(P_5\) on the 
right-hand side.  
Similarly, 
\((\sum_{j=1}^5 P_4^{[j]}\ot\one^{(j)})^2\) on the left-hand side cannot contribute 
to \(P_5\) on the right-hand side.

Thus we can collect terms of weight five on both sides of the equation,
\begin{equation}
\{P_5, \sum_{j=1}^5 P_4^{[j]}\ot\one^{(j)}\} = 6 P_5 \,.
\end{equation}
Tensoring with the identity and multiplying by \(\ket{\phi}\) from the right
leads to
\begin{align}
  \{P_5, \sum_{j=1}^5 P_4^{[j]}\ot\one^{(j)}\} \ot \one^{\ot 2}  \,\ket{\phi} = 6 (P_5 \ot \one^{\ot 2} ) \,\ket{\phi}\,.
\end{align}
However, using the eigenvector relations Eqs.~(\ref{evequation1}, \ref{evequation2}), 
one arrives at a contradiction
\begin{align}
   (2\cdot 5 \cdot 1 + 5\cdot 1 \cdot 2) \,\ket{\phi} = 6\cdot2 \,\ket{\phi}\,.
\end{align}
This ends the proof.
$\hfill \Box$

{\it Upper bound for the number of maximally mixed reductions.---}
Note that in the derivation above not all constraints imposed by the reduced states 
have been taken into account. 
In fact, we only needed a single five-qubit reduced state (say, for definiteness, 
on the qubits $\{1,2,3,4,5\}$) fulfilling the Eqs.~(\ref{projequation}, \ref{evequation2}), 
whose three-body reduced density matrices are all maximally mixed 
[this was needed for Eq.~(\ref{blochfiveequation})]. In addition, the five four-qubit 
reduced density matrices corresponding to the possible subsets of $\{1,2,3,4,5\}$ have 
to obey Eq.~(\ref{evequation1}).

Thus one can try to answer a relaxation of the original
question: Given a seven-qubit state whose two-party reduced states are
all maximally mixed, how many of its three-party reduced states can then
be maximally mixed?
Consider a pure seven-qubit state  where all two-body marginals are 
maximally mixed. This implies that any of the $\binom{7}{5}=21$ possible $\varrho_{(5)}$ 
obeys Eqs.~(\ref{projequation}, \ref{evequation2}). 
There are $\binom{7}{3}=35$ 
possible $\varrho_{(3)}$ and corresponding $\varrho_{(4)}$. If a single three-qubit
reduced state $\varrho_{(3)}$ (say, $\{1,2,3\}$ for definiteness) is not maximally 
mixed, then nine of the $\varrho_{(5)}$ cannot be used for the proof anymore:
First, for six five-qubit subsets (namely, $\{1,2,3,4,5\}, \dots , \{1,2,3,6,7\}$) not
all three-qubit density matrices are maximally mixed, implying that Eq.~(\ref{blochfiveequation})
is not valid. Furthermore, for three five-qubit subsets (namely, $\{1,4,5,6,7 \}$,
$\{2,4,5,6,7\}$, and $\{3,4,5,6,7\}$) not all reduced four-qubit subsets obey 
Eq.~(\ref{evequation1}). It follows that if {\it two} three-qubit reduced states are not 
maximally mixed then at least $21 - 2 \cdot 9 = 3$ five-qubit sets still obey the 
conditions required for the proof. So we can summarize:

\noindent
{\bf Observation 3.}
{
\it Let $\ket{\phi}$ be a pure state of seven qubits, where all two-body reduced
density matrices are maximally mixed. Then, maximally $32$ of the $35$ three-body 
density matrices can be maximally mixed. There exist seven-qubit states for which 
this bound is reached. 
} 

\begin{figure}[t]
\includegraphics[width=0.8 \columnwidth]{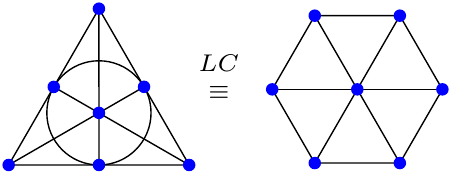}
\caption{
The graph of the Fano (or seven-point) plane on the left, 
which can be transformed by local complementation 
(corresponding to local Clifford gates) to 
the wheel graph displayed on the right.
The Fano plane plays a role in classical error correction, describing both a 
balanced block design as well as an error correcting code \cite{Hill_coding_theory1983}.
The corresponding graph state saturates the bound of Observation~3.  
The states are locally equivalent to the graph state depicted
in Figs.~$1$ in~\cite{Grassl2002}, to No.~$44$ in Table~V from 
Ref.~\cite{Hein2006}, and to the states of Eq.~($11$) in Ref.~\cite{Zha2011}
and of Eq.~($26$) in Ref.~\cite{Goyeneche2014}.
}
\label{fig:fano_plane}
\end{figure}

We note that the existence of states where 32 of the three-body density matrices 
are maximally mixed was shown before: Refs.~\cite{Zha2011, Goyeneche2014} presented such states, 
which are, up to local unitary transformation, a graph state occurring in 
Refs.~\cite{Grassl2002, Hein2006}. As a graph state, the state can be described by the graphs
in Fig.~\ref{fig:fano_plane}. It can be constructed from the graph
as follows: 
Each vertex in a graph corresponds to a qubit. One prepares all the qubits in 
the state $\ket{+}=(\ket{0}+\ket{1})/\sqrt{2}$. Then, for any edge connecting 
the qubits \(j\) and \(k\) one applies a two-qubit phase gate
\begin{equation}
C_{jk} = 
\begin{pmatrix}
1 & 0 & 0 & 0 \\
0 & 1 & 0 & 0 \\
0 & 0 & 1 & 0 \\
0 & 0 & 0 & -1 \\
\end{pmatrix}
\end{equation}
to the initial state. The fact that the marginals of this state have the right 
properties can also directly be checked in the stabilizer formalism, as explained 
in Ref.~\cite{Hein2006}. Finally, we add that there exists an AME state for seven {\it three}-dimensional
systems, which is also a graph state \cite{Helwig2013_graph}.

{\it AME states of \(n\) qubits.---}
The method presented for seven qubits can also 
be applied to the general \(n\)-qubit case. 
There, it can \(exclude\) that an AME state for a given number of qubits exists.
It turns out that the qubit numbers \(n\) for which no contradiction is found (\(n=2,3,5,6\))
are exactly the ones for which AME states are known \cite{Goyeneche2015}.
The proof is presented in the Appendix B \cite{suppmat}.

{\it Conclusion.---}
In summary, we have developed a method based on the 
Bloch representation for characterizing AME states. 
This allowed to rederive most of the known results 
for qubits in a very simple manner, but more importantly, 
it solved the long-standing question whether AME states 
of seven qubits exist or not. Also, the best approximation 
to such a state could be determined. 
For future work, it is very interesting to apply our methods
to the question whether $n$-qubit states exist where all 
$k$-body reduced density are maximally mixed for 
$k <\lfloor n/2\rfloor.$ These are not AME states, but 
they are central for quantum error correction and many 
efforts have been devoted to finding them in the last 
years \cite{Grassl, Feng2015}. We hope that our method can also
contribute to this problem.

We thank 
Sara Di Martino, 
Jens Eisert, 
Christopher Eltschka, 
Mariami Gachechiladze,
Dardo Goyeneche,
Markus Grassl, 
Marcus Huber, 
Christian Majenz,
and Karol {\.Z}yczkowski for fruitful discussions.
This work was supported by 
the Swiss National Science Foundation (Doc.Mobility grant 165024), 
the COST Action MP1209, 
the FQXi Fund (Silicon Valley Community Foundation), 
the DFG, 
the ERC (Consolidator Grant 683107/TempoQ),
the Basque Government grant IT-472-10,
MINECO (Ministry of Science and Innovation of Spain) grants FIS2012-36673-C03-01 and FIS2015-67161-P, 
and the UPV/EHU program UFI 11/55.
Finally, we thank the 
Centro de Ciencias de Benasque Pedro Pascual
and the Cafe Central in Innsbruck for hospitality.

\section{Appendix A}
{From} the projector property, we obtain bounds on AME states. These
originate in work of Rains in the context of quantum codes and were applied 
to AME states by Scott \cite{Rains1998, Scott2004}. 
Let \(\{\Lambda_\alpha\}\) form an orthonormal basis of Hermitian operators for a qudit system
of local dimension \(D\). 
Because of orthonormality,
\(\tr[\Lambda_\alpha \Lambda_\beta] = D \delta_{\alpha\beta}\). 

A \(k\)-body reduced state on parties in \(V\) can then be written as
\begin{equation}
    \varrho_{(k)} 
    = \frac{1}{D^k}(\one + \sum_{\supp(\alpha) \in V} r_{\alpha_1, \dots, \alpha_n} \Lambda_{\alpha_1} \ot \cdots \ot \Lambda_{\alpha_n}) \,.
\end{equation}
Here, the sum runs over appropriate \(\alpha\), specifically, over those
whose corresponding basis terms have nontrivial support only strictly 
within the reduced state under discussion, \( \supp(\alpha) \in V\), 
cf.\ also Eq. \eqref{eq:3qb_trace}.
We recall that any subsystem of an AME state, having size \(k \geq \lfloor~n/2~\rfloor~+~1\), fulfills 
the projector property
\begin{equation}
    \varrho_{(k)}^2 = D^{-(n-k)} \varrho_{(k)}\,.
\end{equation}
Expanding in the Bloch representation and
taking the trace gives 
\begin{equation}
	\tr[ \varrho_{(k)}^2]  
	= \frac{1}{D^k} (1 + \!\!\sum_{\supp(\alpha)\in V} \!\!\!\!\!\! r_{\alpha}^2) = D^{-(n-k)}\,.
\end{equation}
Thus the coefficients \(r_\alpha\) are constrained by
\begin{equation}
	\sum_{\supp(\alpha) \in V} r_{\alpha}^2 = 
	\begin{cases}
	D^{2k-n} -1   > 0 \quad &k > \nhf\,, \\
		        0 \quad &k \leq \nhf\,.
	\end{cases}
\end{equation}
For \(k > \nhf\), the sum is strictly positive, 
because reductions of pure states to size
\(\nhf+1\) can not be proportional to the identity,
as one can see from its Schmidt decomposition.

Let us look at a specific reduced state of size \(\nhf+2\), 
containing \(\nhf+2\) reduced systems of size \(\nhf+1\).
Clearly, all coefficients appearing in the smaller subsystems also appear in the larger subsystem. 

To obtain the bound, we require the coefficients corresponding to weight \(\nhf+2\) alone to be non-negative,
\begin{align}
    \sum_{\substack{ \supp(\alpha) \in V \\ \operatorname{wt} (\alpha) = \alpha + 2}} \!\!\!\! r_{\alpha}^2 
  = &\!\! \sum_{\substack{\supp(\alpha) \in V \\ \operatorname{wt}(\alpha) \leq \nhf+2}} \!\!\!\!\!\!\!\! r_\alpha^2 
  - (\nhf+2) \!\!\!\!\!\!\!\! \sum_{\substack{\supp(\alpha) \in V \\ \operatorname{wt}(\alpha) = \nhf+1}} \!\!\!\!\!\!\!\! r_\alpha^2  \nn\\ 
  \geq & \quad 0\,.
\end{align}
This leads to the conditions
\begin{align}
  (D^{4}-1) - (\nhf + 2)(D^2-1) & \geq 0	\quad\quad n\,\,\rm{even,}	\nn\\
  (D^{3}-1) - (\nhf + 2)(D-1)   & \geq 0	\quad\quad n\,\,\rm{odd,}
\end{align}
which can be recast to the bounds of Refs. \cite{Rains1998, Scott2004},
\begin{equation}
  n \leq 
  \begin{cases}
  	2(D^2-1)  &\quad n\,\,\rm{even,} \\
  	2D(D+1)-1 &\quad n\,\,\rm{odd.}
  \end{cases}
\end{equation}
This ends the proof.

\section{Appendix B}
The general case of determining which \(n\)-qubit AME states can possibly exist 
is detailed here.
It follows the method which was used in the case of seven qubits: We combine 
the projector property of the reduced state of the first \(\nhf +2\) parties with 
the eigenvector relations for the terms \(P_{\nhf+1}\) and \(P_{\nhf+2}\) appearing 
in its expansion.
Collecting terms with either even or odd weight, depending on the case, and applying the parity rule 
will lead to contradictions except in the cases of \(n=2,3,5,6\) qubits.
In the following, we will distinguish four cases, depending on 
\(n\) and \(\nhf\) being even or odd.

\vspace{1em}
{\it Case 1 (\(n\) even, \(\nhf\) even):}
      For \(n\) even, one obtains the two eigenvector relations
      \begin{align}\label{eq:eigenval_even}
	P_{\nhf+1} \ot \one^{\ot (\nhf - 1)} \,\ket{\phi} &= 3 		  \,\ket{\phi} \,,\nn\\
	P_{\nhf+2} \ot \one^{\ot (\nhf - 2)} \,\ket{\phi} &= (9 - 3 \nhf) \,\ket{\phi} \,.
      \end{align}
      Applying the parity rule, we collect terms of odd weight in \(\varrho_{(\nhf+2)}^2\),
      \begin{align}
	    \{ \sum_{j=1}^{\nhf +2} P_{\nhf +1}^{[j]} \ot \one^{(j)}, P_{\nhf +2} \} \,\ket{\phi} \nn\\ 
	    = 14 \sum_{j=1}^{\nhf +2} P_{\nhf +1}^{[j]} \ot \one^{(j)} \,\ket{\phi} \,.
      \end{align}
      This results in a contradiction, as
      \begin{equation}
	    9- 3\nhf \neq 7\,.
      \end{equation}
      Thus qubit AME states do not exist when \(n\) is a multiple of \(4\).

\begin{figure}[t]
\includegraphics[width= \columnwidth]{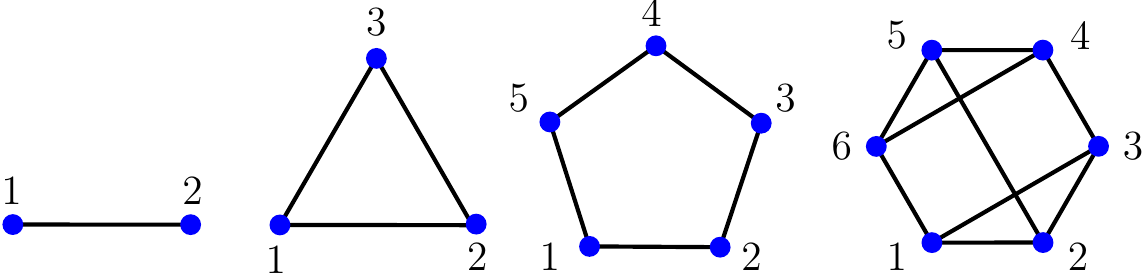}
\caption{
These graphs correspond to AME graph states of two, three, five, and six qubits.}
\label{fig:all_AME}
\end{figure}

\vspace{1em}
{\it Case 2 (\(n\) even, \(\nhf\) odd):}
      The eigenvector relations are as appearing in Case 1, Eq.~\eqref{eq:eigenval_even}. 
      We collect terms of odd weight in \(\varrho_{(\nhf+2)}^2\),
      \begin{align}
	    \{ \sum_{j=1}^{\nhf +2} P_{\nhf +1}^{[j]} \ot \one^{(j)}, P_{\nhf +2} \} \ket{\phi} \,,\nn\\
	    = 14 P_{ \nhf +2}
      \end{align}
      If \(P_{\nhf +2} \ket{\phi} \neq 0\), we obtain a contradiction because
      \begin{equation}
      	(\nhf+2)\cdot 3 \neq 7\,.
      \end{equation}
      Thus \(\one^{\ot (\nhf - 2)} \ot P_{\nhf +2} \ket{\phi} = 0\).
      But from the eigenvector relation in Eq.~\eqref{eq:eigenval_even}
      this can only by possible if \(n=6\). Indeed, for this case an AME graph state
      is known, depicted in Fig.~\ref{fig:all_AME}. Note that the Bell state consisting of 
      only two qubits is too small to be excluded by this method.

\vspace{1em}
{\it Case 3 (\(n\) odd, \(\nhf\) even):}
      For \(n\) odd, one obtains the two eigenvector relations
      \begin{align} \label{eq:eigenval_odd}
	P_{\nhf+1} \ot \one^{\ot (\nhf - 1)} \,\ket{\phi} &= 		\,\ket{\phi} \,,\nn\\
	P_{\nhf+2} \ot \one^{\ot (\nhf - 2)} \,\ket{\phi} &= (5 - \nhf) \,\ket{\phi} \,.
      \end{align}
      We collect terms of odd weight,
      \begin{align}
	    \{ \sum_{j=1}^{\nhf +2}  P_{\nhf +1}^{[j]} \ot \one^{(j)}, P_{\nhf +2} \} \ket{\phi} \nn\\ 
	    = 6 \sum_{j=1}^{\nhf +2} P_{\nhf +1}^{[j]} \ot \one^{(j)} \ket{\phi} \,.
      \end{align}
      Thus one requires
      \begin{align}
	    (5 - \nhf) \ket{\phi} = 3 \ket{\phi} \,,
      \end{align}
      whose only solution is \(n=5\). The corresponding AME state 
      is the five-qubit ring-cluster state, depicted in Fig.~\ref{fig:all_AME}.

\vspace{1em}
{\it Case 4 (\(n\) odd, \(\nhf\) odd):}
      This final case is slightly more involved, but the method
      ultimately succeeds on a larger reduced state of size \(\nhf+4\).
      The eigenvector relations are as appearing in Case~3, Eq.~\eqref{eq:eigenval_odd}.
      We collect terms of odd weight,
      \begin{align}
	    \{ \sum_{j=1}^{\nhf +2}  P_{\nhf +1}^j \ot \one^{(j)}, P_{\nhf +2} \} \,\ket{\phi} \nn\\ 
	    = 6 P_{\nhf +2} \,\ket{\phi} \,.
      \end{align}
      If \(P_{\nhf +2} \ket{\phi} \neq 0\), it follows that
      \begin{align}
      	\nhf + 2 = 3\,.
      \end{align}
      The only solution is \(n=3\), corresponding to the GHZ state.
      If however \(P_{\nhf +2} \ket{\phi} = 0\), that is \(n=11\),
      we have to make use of further eigenvector relations.
      \begin{align}
	P_{6} \ot \one^{\ot 5} \,\ket{\phi} &= 1 \,\ket{\phi}   \,,\nn\\
	P_{7} \ot \one^{\ot 4} \,\ket{\phi} &= 0 \,\ket{\phi}   \,,\nn\\
	P_{8} \ot \one^{\ot 3} \,\ket{\phi} &= 3 \,\ket{\phi}   \,,\nn\\
	P_{9} \ot \one^{\ot 2} \,\ket{\phi} &= 16 \,\ket{\phi}  \,.
      \end{align}
      We require \(\varrho_{(9)}^2 = 2^{-2} \varrho_{(9)}\) and collect terms of odd weight
      \begin{align}
	(\{ \sum_{j=1}^{\binom{9}{6}} P_6^{[j]} \ot \one^{\ot 3} , P_9 \} + \{ \sum_{j=1}^{9} P_8^{[j]} \ot \one , P_9 \} ) \,\ket{\phi} \nn\\
	= 126 P_9 \,\ket{\phi} \,.
      \end{align}
      This also leads to a contradiction in the case of \(n=11\),
      \begin{equation}
      	\binom{9}{6} + 9\cdot 3 \neq 63\,.
      \end{equation}
      Therefore, no AME qubit state with both \(n\) and \(\nhf\) being odd exists.

We summarize: the only qubit AME states which are not excluded to exist by this method 
are the cases of two, three, five, and six parties, all of which are known.
Their graph state representations are shown in 
Fig.~\ref{fig:all_AME}.


\begin{thebibliography}{99}
 
\bibitem{Gisin1998}
  N.\ Gisin and H.\ Bechmann-Pasquinucci,
  Phys. Lett. A {\bf 246}, 1  (1998).

\bibitem{Higuchi2000}
  A.\ Higuchi and A.\ Sudbery,
  Phys. Lett. A {\bf 273}, 213 (2000).
  
\bibitem{Scott2004}
  A.\ J.\ Scott,
  Phys. Rev. A {\bf 69}, 052330 (2004).

\bibitem{Brown2005}
  I. D. K. Brown, S. Stepney, A. Sudbery, and S. L. Braunstein,
  J. Phys. A: Math. Gen {\bf 38}, 1119 (2005).

\bibitem{Borras2007}
  A. Borras, A. R. Plastino, J. Batle, C. Zander, M. Casas, and A. Plastino,
  J. Phys. A: Math. Theor. {\bf 44}, 13407 (2007).

\bibitem{Facchi2008}
   P. Facchi, G. Florio, G. Parisi, and S. Pascazio,
   Phys. Rev. A {\bf 77}, 060304 (2008).
  
\bibitem{Facchi2010}
  P. Facchi, G. Florio, U. Marzolino, S. Pascazio, and G. Parisi, 
  J.\ Phys.\ A: Math.\ Theor.\ {\bf 43},  225303  (2010).
 
 \bibitem{Gour2010}
  G. Gour and N. R. Wallach,
  J. Math. Phys. {\bf 51}, 112201 (2010).
  
 \bibitem{Zha2011}
  X.-W. Zha, H.-Y. Song, J.-X. Qi, D. W., and Q. Lan, 
  J. Phys. A: Math. Theor. {\bf 45}, 255302 (2012).

\bibitem{Helwig2012}
  W. Helwig, W. Cui, J. I. Latorre, A. Riera, and H.-K. Lo,
  Phys. Rev. A {\bf 86}, 052335 (2012).  

\bibitem{Arnoud2013}
  L. Arnaud and N. J. Cerf,
  Phys. Rev. A {\bf 87}, 012319 (2013).  

\bibitem{Helwig2013}
  W. Helwig and W. Cui,
  arXiv:1306.2536. 
  
\bibitem{Helwig2013_graph}  
 W. Helwig, 
 arXiv:1306.2879.
  
\bibitem{Kloeckl2015}
  C. Kl\"ockl and M. Huber, 
  Phys. Rev. A. {\bf 91}, 042339 (2015).

 \bibitem{Goyeneche2015}
  D. Goyeneche, D. Alsina, J. I. Latorre, A. Riera, and K. {\.Z}yczkowski, 
  Phys. Rev. A {\bf 92}, 032316 (2015).
  
\bibitem{Zyczkowski2016}
  M Enr\'{i}quez, I. Wintrowicz and K. {\.Z}yczkowski,
  J. Phys. Conf. {\bf 1}, 012003 (2016).
 
\bibitem{Chen2016}
  L. Chen and D. L. Zhou, 
  Scientific Reports {\bf 6}, 27135 (2016).
  
\bibitem{Bernal2016}  
  A. Bernal,
  arXiv:1603.06082.
  
\bibitem{Hein2006}
  M. Hein, J. Eisert, and H.J. Briegel,
  Phys. Rev. A {\bf 69}, 062311 (2004).
    
\bibitem{Rains1998}
  E. M. Rains,
  IEEE Trans. Inf. Theory {\bf 44}, 1388 (1998).

\bibitem{Rains1999}
  E. M. Rains,
  IEEE Trans. Inf. Theory {\bf 45}, 2361 (1999).
  
\bibitem{Nebe2006}
  G. Nebe, E. M. Rains, and N. J. A. Sloane,
  {\it Self-Dual Codes and Invariant Theory},
  Springer Berlin-Heidelberg (2006).

\bibitem{Goyeneche2014}
  D.\ Goyeneche and K.\ {\.Z}yczkowski,
  Phys.\ Rev.\ A {\bf 90}, 022316 (2014).

\bibitem{Grassl2015}
  M.\ Grassl and M.\ Rötteler,
  arxiv:1502.05267

\bibitem{Calderbank1998}
  A.\ R.\ Calderbank, E.\ M.\ Rains, P.\ W.\ Shor, and N.\ J.\ A.\ Sloane,
  IEEE Trans. Inf. Theory {\bf 44}, 1369 (1998)

\bibitem{open_cases}
  Up to \(n=30\), the existence of one-dimensional non-additive codes with parameters
  \( (( 13,1,  6))_2 \), 
  \( (( 19,1,  8))_2 \), and
  \( (( 25,1, 10))_2 \) is still unresolved. See Table~$13.3$ in \cite{Nebe2006}.

\bibitem{Fano1957}
  U. Fano, Rev.\ Mod.\ Phys.\ {\bf 19} 74 (1957).
   
\bibitem{Eltschka2015}
  C.\ Eltschka and J.\ Siewert,
  Phys.\ Rev.\ Lett.\ {\bf 114}, 140402 (2015). 
  
\bibitem{Grassl2002}
  M.\ Grassl, A.\ Klappenecker, and M.\ Rötteler,
  Proc. IEEE Int. Symp. Inf. Theory, 45 (2002),
  quant-ph/0703112.
  
\bibitem{suppmat}
  The Appendix can be found in the supplemental material.

\bibitem{Hill_coding_theory1983}
  R. Hill,
  {\it A First Course in Coding Theory}, 
  Oxford University Press (1983).
  
\bibitem{Grassl} For a current overview on this
  problem see M. Grassl, {\it Bounds on the minimum distance of 
  linear codes and quantum codes}, available at {\tt www.codetables.de}.

\bibitem{Feng2015}
  K. Feng, L. Jin, C. Xing, and C. Yuan,
  arXiv:1511.07992.

\end{thebibliography}
\end{document}